\documentclass[final,1p,times]{elsarticle} 
\usepackage{graphicx} 
\usepackage{amssymb} 
\usepackage{amsthm} 
\usepackage{lineno} 

\journal{Nuclear Physics A} 
\begin{document} 

\begin{frontmatter} 


\title{Jet Observables of Parton Energy Loss \\ in High-Energy Nuclear
Collisions}

\author{ Ben-Wei Zhang$^{a,b}$ }

\address[a]{Los Alamos National Laboratory, Theoretical Division,
MS B238, Los Alamos, NM 87545, USA}
\address[b]{Key Laboratory of Quark $\&$ Lepton Physics (Hua-Zhong Normal
University), Ministry of Education, China}

\begin{abstract} 
While strong attenuation of single particle production and particle 
correlations has provided convincing evidence for large parton energy 
loss in the QGP, its application to jet tomography has inherent limitations 
due to the inclusive nature of the measurements. Generalization of this 
suppression to  full jet observables leads to an unbiased, more differential 
and thus powerful approach to determining the characteristics of the hot 
QCD medium created in high-energy nuclear collisions. In this article we 
report on recent theoretical progress in calculating jet shapes and the
related jet cross sections in the presence of QGP-induced  parton energy 
loss. (i) A theoretical model of intra-jet energy flow in heavy-ion 
collisions is discussed.  (ii) Realistic numerical simulations demonstrate 
the nuclear modification factor  $R_{AA}(p_T)$  evolves continuously 
with the jet cone size $R^{\max}$ or the acceptance cut $\omega_{\min}$ -  
a novel feature of jet quenching. The anticipated broadening of jets is  
subtle and most readily manifested in the periphery of the cone for 
smaller cone radii.

\end{abstract} 

\end{frontmatter} 


\section{Introduction}
\label{Intro}

When a fast quark or gluon traverses a hot/dense nuclear medium, it may undergo
multiple scattering with other partons in the medium and lose a large amount of
its energy via induced gluon bremsstrahlung~\cite{GVWZ}. 
This jet quenching mechanism has been used to successfully  explain the
strong suppression of the hadron spectra at large transverse momentum
observed in nucleus-nucleus collisions at the Relativistic Heavy
Ion Collider (RHIC). However, at present, most measurements of hard 
processes are limited to inclusive hadron (or photon) production and
di-hadron (or gamma-hadron) correlations, which are only the leading fragments 
of a jet. Thus their measurement may suffer from geometric biases \cite{Salur}. 
With the upgrades at the RHIC experimental facilities and the new opportunities 
provided by LHC,  much more differential studies of parton energy loss 
in nuclei will become available - the ability  to investigate the full structures 
of a jet in relativistic heavy-ion collisions  \cite{Ploskon,Grau}.
In this article we review our recent theoretical progress on 
calculating jet shapes and the related jet cross sections in reactions with 
ultra-relativistic nuclei \cite{Vitev:2008rz},
which become feasible as a new, differential and accurate test of the underlying  
QCD theory. Our theoretical approach to understanding the jet shapes in the vacuum 
as well as the  medium-induced jet shapes with experimental acceptance cut 
will be discussed. We will also show numerical simulations and their implications
for the current heavy-ion program.

\section{Jet shapes in p+p collisions}
The jet shape, related to the intra-jet energy flow, is one of the most common ways 
of resolving the internal jet structure. The usual definition of an ``integral jet     
shape''  reads \cite{Seymour:1997kj}:
\begin{equation}
 \Psi_{\rm {int}}(r;R) = \frac{\sum_i (E_T)_i \Theta (r-R_{i, \rm {jet}})}
{\sum_i  (E_T)_i  \Theta(R-R_{i,\rm {jet}})} \, .
\end{equation}
Here $r,R$ are Lorentz-invariant opening angles, $R_{i,{\rm jet}} =                       
\sqrt{(\eta_i-\eta_{\rm {jet}})^2 + (\phi_i-\phi_{\rm {jet}})^2}$ in 
a cone algorithm, and $i$ represents
a sum over all particles in this jet. With the above definition we have
$\Psi_{\rm {int}}(R;R) = 1$. A differential jet shape is the defined as follows:
\begin{equation}
 \psi(r;R) = \frac{d\Psi_{\rm {int}}(r;R)}{dr}  \;,
\end{equation}
and $\psi(r;R)dr$ gives the fraction of all energy within a cone with the size $R$ 
around the jet axis that is within an annulus of radius $r$ and width $dr$, centred
on the jet axis.

\begin{figure*}[t!]
\includegraphics[width=2.2in,height=2.6in,angle=0]{CDF_comparison.eps} \hspace*{1cm}
\includegraphics[width=2.7in,height=2.3in,angle=0]{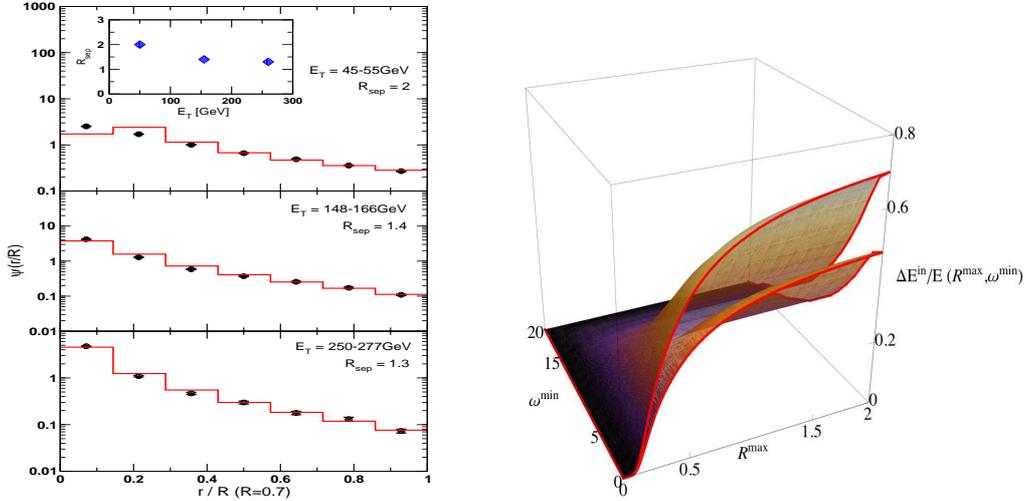}  
\caption[]{(Left panel) Comparison of numerical results
from our theoretical calculation to experimental data on
differential jet shapes at $\sqrt{s}=1960$~GeV by
CDF II~\cite{Acosta:2005ix}. (Right panel) 3D plot for the ratio of the energy that a
partons loses inside a jet cone of opening angle $R^{\max}$  with  $\omega >
\omega^{\min}$ to the total parton energy with $E_{\rm jet}=20$~GeV in $b=3$~fm Pb+Pb
collisions at LHC.}
\label{CDF-dE}
\end{figure*}

We follow an analytical approach by Seymour \cite{Seymour:1997kj}, generalize it
to include finite experimental acceptance cut effect, and  find \cite{Vitev:2008rz}:
\begin{eqnarray}
\psi(r)=\psi_{\texttt{coll}}(r)\left( P(r)-1\right) +
\psi_{\texttt{LO}}(r) + \psi_{i,\texttt{LO}}(r)
+ \psi_{\texttt{PC}}(r)
+ \psi_{i,\texttt{PC}}(r) \; .
\label{totpsi}
\end{eqnarray}
On the right-hand-side of Eq.~(\ref{totpsi}) the first term represents
the contribution from  Sudakov resummation  with  subtraction  to avoid double
counting; the second and  third terms give the leading-order contributions in the 
final-state and the initial-state splitting, respectively; the last two 
terms come from  power corrections when integrating  over the Landau pole. 
In the left panel of Fig.~\ref{CDF-dE} we show the comparison of
our theoretical results for jet shapes to the CDF II data. It can be seen that 
the pQCD calculation yields a good description of the jet shapes measured at 
the CDF, and thus provides a reliable baseline in $p+p$ collisions for comparison
to the full in-medium jet shape in heavy-energy nuclear reactions.

\section{Energy distribution due to medium-induced gluon radiation}
When an energetic parton propagates inside the QGP, it will lose energy
via induced gluon radiation. This will give rise to additional contributions
to jet shapes. In this study, we adopt the GLV formalism to calculate the
in-medium jet shapes~\cite{GLV,Vitev:2008vk}. An important feature 
of the induced  final-state bremsstrahlung in the deep LPM regime   
is that there is no collinear divergence~\cite{Vitev:2008vk} and, thus, 
no resummation is needed when $r \rightarrow 0$. In the GLV formalism, the 
intensity spectrum due to final-state gluon
radiation can be written as \cite{Vitev:2008rz}:
\begin{eqnarray}
k^+ \frac{dN^g(FS)}{dk^+ d^2 {\bf k} } &=& \frac{C_R \alpha_s}{\pi^2}
\sum_{n=1}^{\infty}  \left[ \prod_{i = 1}^n  \int
\frac{d \Delta z_i}{\lambda_g(z_i)}  \right]
\left[ \prod_{j=1}^n \int d^2 {\bf q}_j \left( \frac{1}{\sigma_{el}
(z_j)}
\frac{d \sigma_{el}(z_j) }{d^2 {\bf q}_j}
-  \delta^2 ({\bf q}_j) \right)    \right] \nonumber \\
&\times& \left[ -2\,{\bf C}_{(1, \cdots ,n)} \cdot
\sum_{m=1}^n {\bf B}_{(m+1, \cdots ,n)(m, \cdots, n)}
\left( \cos \left (
\, \sum_{k=2}^m \omega_{(k,\cdots,n)} \Delta z_k \right)
-   \cos \left (\, \sum_{k=1}^m \omega_{(k,\cdots,n)}
\Delta z_k \right) \right)\; \right]  \;.  \qquad 
\nonumber
\end{eqnarray}
With growing jet cone radius $R_{\max}$ more of the lost energy, carried away
by radiated gluons, will fall back again inside the cone; conversely with larger
acceptance cut $\omega^{\min}$, a larger energy fraction will not be measured. 
The right panel of Fig.~\ref{CDF-dE} illustrates
our numerical results for the fractional energy loss inside
the jet cone as a function of the cone size $R_{\max}$ and the experimental 
acceptance cut $\omega^{\min}$, defined as:
\begin{eqnarray}
  \frac{\Delta E^{in}}{E}(R^{\max},\omega^{\min})
= \frac{1}{E} \int_{\omega^{\min}}^E d\omega
  \int_0^{R^{\max} } dr  \frac{dI^g}{d\omega dr} (\omega,r) \; .
\label{def:out}
\end{eqnarray}


\section{Jet tomography in high-energy nuclear collisions}
\begin{figure*}[t!]
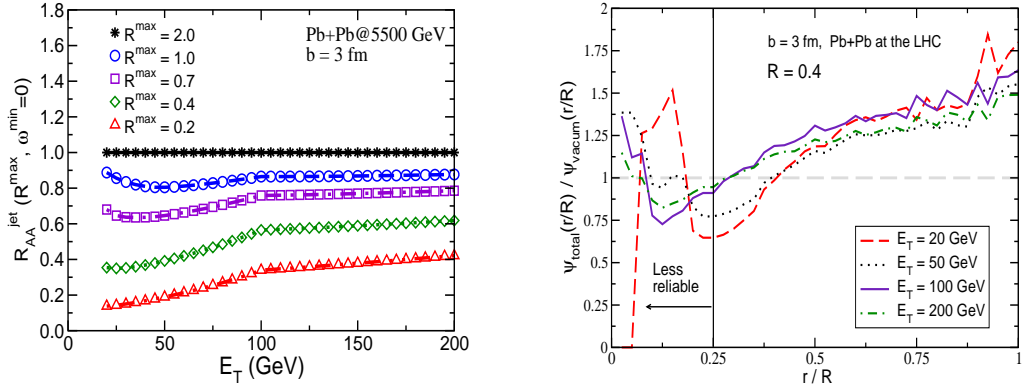

\includegraphics[width=2.4in,height=2.in,angle=0]{Raa-R-b3-jet.eps} \hspace*{1cm}
\includegraphics[width=2.4in,height=2.in,angle=0]{Rpsi-R0.4-W0-jet-N.eps}  
\caption[]{(Left panel) Nuclear modification factor  
$R_{AA}^{\rm jet}(R^{\max},\omega^{\min})$ as a function of $E_T$ 
for different jet cone radii 
$R^{\max}$. (Right panel) The ratios of the total jet shape in 
high-energy nuclear collisions to the jet 
shape in vacuum with cone radius 
$R=0.4$. Results are for Pb+Pb collision at $\sqrt{s}=5.5$~TeV with $b=3$~fm.}
\label{Raa-jet}
\end{figure*}
Taking into account the contributions to the jet shape from vacuum splitting 
and  the medium-induced bremsstrahlung, we can obtain the full jet shape in 
high-energy nuclear collisions as:
\begin{eqnarray}
\psi_{\rm tot.}\left({r}/{R}\right) &=&  
\frac{1}{\rm Norm}  \int_{\epsilon=0}^1 
d\epsilon \; \sum_{q,g} P_{q,g}(\epsilon) 
\frac{1}{ (1 - (1-f_{q,g}) \cdot \epsilon)^3} \nonumber \\
&&\times
\frac{\sigma^{NN}_{q,g}(R,\omega^{\min})} {d^2E^\prime_Tdy} 
\Big[ (1- \epsilon) \; 
\psi_{\rm vac.}^{q,g}\left({r}/{R}\right) 
 +  \, f_{q,g}\cdot \epsilon \; 
\psi_{\rm med.}^{q,g}\left(r/R\right) \Big] \; . \qquad
\label{Psi:total}
\end{eqnarray} 
In Eq.~(\ref{Psi:total}) $f =
 {\Delta E_{\rm rad}\left\{ (0,R);(\omega^{\min},E) \right\} }/
{\Delta E_{\rm rad} \left\{ (0,R^\infty);(0,E) \right\} } $ gives
the fraction of the lost energy that falls within the jet cone, $r < R$, 
and carried by gluons of $\omega > \omega^{\min}$ relative to the 
total parton energy loss without the above kinematic constraints. Furthermore, 
$\epsilon$ is the total fractional energy loss and $P(\epsilon)$ represents the related 
probability distribution.

We can also calculate the related jet cross section with parton energy
loss in $A+A$ collisions \cite{Vitev:2008rz}  and generalize the nuclear modification 
factor of leading hadrons to that of jets as follows:
\begin{equation}
R_{AA}^{{\rm jet}}(E_T; R^{\max},\omega^{\min})  =
\frac{ \frac{d\sigma^{AA}(E_T;R^{\max},\omega^{\min})}{dy d^2 E_T} }
{ \langle  N_{\rm bin}  \rangle
\frac{d\sigma^{pp}(E_T;R^{\max},\omega^{\min})}{dy d^2 E_T}  } \; .
\label{RAAjet}
\end{equation}
Based on the analytic pQCD  model we are now ready to perform numerical simulations
and several selected results are shown in Fig.~\ref{Raa-jet}. 
We see that with altering  cone radius $R_{AA}^{{\rm jet}}(E_T; R^{\max},\omega^{\min})$ 
changes continuously and reaches unity for large radii  ($R^{\max} =2.0$) when
all of the lost energy falls back inside the cone. This is in
stark contrast to a single $R_{AA}$ curve for inclusive hadron production observed at 
RHIC. Therefore, measurements of the suppression of jet cross sections for different $R^{\rm max}$
will provide an independent and much more accurate way  to determine the characteristics 
of parton energy loss in the QGP. 
The right panel of Fig.~\ref{Raa-jet} illustrates that the QGP broadening effects are
manifest in the tails of the energy flow distribution and the enhancement factor due to
medium-induced jet shapes can be about $1.75$ when $r/R \sim 1$ with $R=0.4$. 
 
\section*{Acknowledgments} 
We thank I. Vitev, S. Wicks, M. H. Seymour, H. Caines, N. Grau, H. Takai, S. Salur and M. Ploskon 
for many helpful discussions. This research is supported by the US Department of Energy, Office 
of Science, under Contract No. DE-AC52-06NA25396 and in part by the LDRD program 
at LANL, the MOE of China under Project No. IRT0624 and the NNSF of China.


\begin{thebibliography}{00} 

\bibitem{GVWZ}
  M.~Gyulassy, I.~Vitev, X.~N.~Wang and B.~W.~Zhang,
  arXiv:nucl-th/0302077.

\bibitem{Salur}
  S.~Salur  [for the STAR Collaboration],
  arXiv:0907.4536 [nucl-ex].


\bibitem{Ploskon}
 M.~Ploskon [for the STAR Collaboration], arXiv:0908.1799 [nucl-ex].

\bibitem{Grau}
N. Grau [for the ATLAS collaboration], arXiv:0907.4944 [nucl-ex].

\bibitem{Vitev:2008rz}
  I.~Vitev, S.~Wicks and B.~W.~Zhang,
  JHEP {\bf 0811} (2008) 093
  [arXiv:0810.2807 [hep-ph]]; \\
I.~Vitev, B.~W.~Zhang and S.~Wicks,
  Eur.\ Phys.\ J.\  C {\bf 62}, 139 (2009)
  [arXiv:0810.3052 [hep-ph]].


\bibitem{Seymour:1997kj}
  M.~H.~Seymour,
  Nucl.\ Phys.\  B {\bf 513}, 269 (1998);

\bibitem{Acosta:2005ix}
  D.~E.~Acosta {\it et al.}  [CDF Collaboration],
  Phys.\ Rev.\  D {\bf 71}, 112002 (2005)
  [arXiv:hep-ex/0505013].

\bibitem{GLV}
  M.~Gyulassy, P.~Levai and I.~Vitev,
  Phys.\ Rev.\ Lett.\  {\bf 85}, 5535 (2000);
  I.~Vitev,
  Phys.\ Rev.\  C {\bf 75}, 064906 (2007).

\bibitem{Vitev:2008vk}
  I.~Vitev and B.~W.~Zhang,
  Phys.\ Lett.\  B {\bf 669} (2008) 337; 
I.~Vitev,
  Phys.\ Lett.\  B {\bf 630}, 78 (2005).







\end{thebibliography}
\end{document}